\documentclass[aps,twocolumn,showpacs,preprintnumbers,nofootinbib,prd,superscriptaddress,groupedaddress,10pt]{revtex4-1}
\usepackage[utf8]{inputenc}
\makeatletter
\def\l@subsubsection#1#2{}
\def\l@subsubsubsection#1#2{}
\makeatother
\usepackage{amsmath}
\usepackage{graphicx,amssymb,amsmath,amsthm,amsfonts,epsfig,epsf}
\usepackage[usenames,dvipsnames,svgnames,table]{xcolor}
\usepackage{epstopdf}
\definecolor{darkred}{rgb}{0.5,0,0}
\usepackage{aas_macros}
\usepackage{bm}
\usepackage{dcolumn}
\usepackage[utf8]{inputenc}
\usepackage{latexsym}
\usepackage{rotating}
\usepackage{longtable}

\usepackage{enumerate}
\usepackage{tensor,multirow}
\usepackage{mathtools}
\usepackage{url}
\usepackage[linktocpage,hidelinks]{hyperref}
\newcommand{\mndd}{_{\mu\nu}}
\newcommand{\mnuu}{^{\mu\nu}}

\newcommand{\rgb}{\mathcal{R}_{\rm GB}}
\newcommand{\til}{~}
\def\nn{\nonumber}

\begin{document}

\title{Spin-induced scalarization and magnetic fields}

\author{
Lorenzo Annulli$^{1}$,  %
Carlos A. R. Herdeiro$^{1}$
and
Eugen Radu$^{1}$
}

\affiliation{${^1}$ Departamento de Matem\'{a}tica da Universidade de Aveiro and Centre for Research and Development in Mathematics and Applications (CIDMA), Campus de Santiago, 3810-183 Aveiro, Portugal}

\begin{abstract}
In the presence of certain non-minimal couplings between a scalar field and the Gauss-Bonnet curvature invariant, Kerr black holes can scalarize, as long as they are spinning fast enough. This provides a distinctive violation of the Kerr hypothesis, occurring only for some high spin range. In this paper we assess if strong magnetic fields, that may exist in the vicinity of astrophysical black holes, could facilitate this distinctive effect, by bringing down the spin threshold for scalarization. This inquiry is motivated by the fact that self-gravitating magnetic fields, by themselves, can also promote ``spin-induced'' scalarization.  Nonetheless, we show that in the \textit{vicinity of the horizon} the effect of the magnetic field $B$ on a black hole of mass $M$, up to $BM\lesssim 1$, works \textit{against} spin-induced scalarization, requiring a larger dimensionless spin $j$ from the black hole. A geometric interpretation for this result is suggested, in terms of the effects of rotation $vs.$ magnetic fields on the horizon geometry.
%We show that, in theories with non-trivial couplings between scalar fields and higher-order curvature invariants, the necessary condition for charged and spinning black holes to be prone to tachyonic instabilities is satisfied for higher values of the spin, if magnetic fields are involved.
%, compared to the case in which magnetic fields are absent
%This perhaps counter-intuitive result might be explained in light of the inter-playing roles that the magnetic field and the spin have on the horizon geometry.
\end{abstract}

\maketitle

%\tableofcontents

%%%%%%%%%%%%%%%%%%%%%%%%%%%%%%%%%%%%
\section{Introduction}
%%%%%%%%%%%%%%%%%%%%%%%%%%%%%%%%%%%%
The prospect of testing the \textit{Kerr hypothesis} is now more realistic than ever. Are all black holes (BHs) in the astrophysical mass range ($\sim$ 1  to $10^{10}$ $M_\odot$), and for all spins really well described by the Kerr metric~\cite{Kerr:1963ud}? The unprecedented access to strong gravity data, both through gravitational waves\til\cite{Abbott:2016blz,Abbott:2016nmj,Abbott:2017vtc,Abbott:2017oio,Abbott:2020tfl,LIGOScientific:2021usb} and from electromagnetic observations~\cite{GRAVITY:2020gka,EventHorizonTelescope:2019dse} will provide hints, or even clear answers, about this central question in strong gravity. 

Amongst the Kerr hypothesis violating models, a sub-class still accommodates the Kerr solution of vacuum General Relativity (GR), but questions the \textit{universality} of the hypothesis. Such models typically introduce new scales; these define a mass/spin range in which BHs can become non-Kerr, whereas Kerr BHs remain as the solution of the model in a complementary range.

A concrete illustration of this sub-class of models occurs via the mechanism of \textit{spontaneous scalarization}. Originally proposed in scalar-tensor theories and for neutron stars\til\cite{Damour:1993hw}, the spontaneous scalarization of vacuum GR BHs occurs in extended scalar-tensor models, wherein a real scalar field with a canonical kinetic term, requiring no mass or self-interactions, minimally couples to the Gauss-Bonnet (GB) quadratic curvature invariant, via a coupling $\eta f(\phi)$, where $\eta$ is a coupling constant and $f(\phi)>0$ a coupling function. We are then led to consider electrovacuum GR augmented by such a scalar field and such a GB curvature correction: an Einstein-Maxwell-scalar-GB (EMsGB) model, $cf.$ action~\eqref{eq:EMsGB_action} below. 

Depending on the sign of the coupling function, $\eta$, the scalarization has different triggers. For $\eta>0$ Schwarzschild BHs with mass $M$ scalarize, as long as $M/\sqrt{\eta}$ is small enough\til\cite{Silva:2017uqg,Doneva:2017bvd,Antoniou:2017acq}. Adding spin to the Schwarzschild BH actually quenches the effects of the scalarization~\cite{Cunha:2019dwb,Collodel:2019kkx}, albeit alleviating slightly the constraint on $M/\sqrt{\eta}$. In this case scalarization, which is due to a tachyonic instability, is promoted by spacetime regions wherein the GB curvature invariant is (sufficiently) \textit{positive}. Following~\cite{Herdeiro:2021vjo} we call this GB$^+$ scalarization.

For $\eta<0$, on the other hand, the tachyonic instability is promoted by spacetime regions where the GB curvature invariant is \textit{negative}, hereafter dubbed GB$^-$ scalarization. For vacuum GR BHs - $i.e.$ the Kerr family - this occurs for dimensionless spin $j>0.5$. Since the negative GB patches occur in the strong gravity region neighbouring the horizon, they trigger scalarization immediately as they appear. Thus Kerr BHs with $j>0.5$ undergo scalarization in these models, which was dubbed ``spin-induced" scalarization~\cite{Dima:2020yac}. Spin-induced scalarization can thus be regarded as a concrete example of GB$^-$ scalarization. The corresponding scalarized BHs were constructed in\til\cite{Herdeiro:2020wei,Berti:2020kgk}.

GB$^-$ scalarization can also be triggered by other sources rather than BH spin. Of particular interest for us is the case of self-gravitating magnetic fields. In electrovacuum GR, these are described by the Melvin magnetic Universe\til\cite{Melvin:1963qx}. It is known that such universes have regions with a negative GB invariant\til\cite{Brihaye:2021jop} and can thus promote GB$^-$ scalarization. One may then ask if the presence of a strong magnetic field in the vicinity of a spinning BH may facilitate GB$^-$ (\textit{aka}, in this context, spin-induced) scalarization, promoting it for $j<0.5$. The goal of this paper is to answer this question, showing that this is \textit{not} the case.

Astrophysical BHs are thought to be neutrally charged, due to, $e.g.$ quantum discharge\til\cite{Gibbons:1975kk} and electron-positron pair production\til\cite{Blandford:1977ds}. The inclusion of magnetic fields in BH environments is, nonetheless, well motivated. There is in fact observational evidence of astrophysical magnetic fields; elongated sources present in radio-relics\til\cite{2017A&A...600A..18K} -- diffuse radio sources in galaxy clusters\til\cite{Kempner:2003eh} -- are an example. Furthermore, there are known observations supporting the existence of BHs immersed in magnetic backgrounds, as the case of the presence of the magnetar SGR J$1745-29$ in the vicinity of Sagittarius A$^*$\til\cite{Mori:2013yda,Kennea:2013dfa,Eatough:2013nva,Olausen:2013bpa}, or the strong magnetic fields in the proximity of the event horizon of M87$^*$\til\cite{EventHorizonTelescope:2021srq}. 
%Furthermore, from a theoretical perspective, it is also known that geodesic motion of charged particles around BHs is crucially influenced by magnetic fields of the order of $10^{19}{\rm G}$\til\cite{Frolov:2010mi}. 
%In light of the above, it is fundamental to have BH solutions in closed form that properly account for magnetic fields, both in GR and in modified theories of gravity.

In this work, we investigate the effects on GB$^-$ scalarization due to magnetic fields around spinning BH spacetimes, in the EMsGB model. Our strategy will be to consider spinning BHs in Melvin Universes in electrovacuum, investigating the impact of the magnetic field on the regions where the GB invariant becomes negative. Although Melvin Universes are irrealistic to describe globally astrophysical BHs, due to the non-trivial asymptotics, one may argue that in the vicinity of the BH, where the relevant physics that we are investigating is developing, they capture, approximately, the appropriate physics  in the case of poloidal magnetic fields. 

This paper is organized as follows. In Sec.\til\ref{Theoretical_setup}
 the EMsGB model is described and its relevant equations are shown. In Sec.\til\ref{Magnetized_solutions} we review the relevant electrovacuum GR  solutions describing a BH immersed in magnetic environments. The main analysis and results of this paper are presented in Sec.~\ref{Scalarization_and_magnetized_spacetimes} where we show how and when the necessary condition for ${\rm GB}^-$ scalarization can be satisfied in magnetized BH spacetimes. Finally, conclusions are provided in Sec.\ref{sec:conclusions}.

%%%%%%%%%%%%%%%%%%%%%%%%%%%%%%%%%%%%
\section{Theoretical setup}
\label{Theoretical_setup}
%%%%%%%%%%%%%%%%%%%%%%%%%%%%%%%%%%%%
The action of the EMsGB model is
\begin{equation} \label{eq:EMsGB_action}
\frac{1}{16\pi}\int d^4x\sqrt{-g}\left[R-F\mndd F\mnuu-\frac{1}{2}\left(\nabla \phi\right)^2+\frac{\eta}{4} \mathfrak{f}(\phi)\rgb\right]
\end{equation}
where $F\mndd$ is the usual Faraday tensor, defined by means of the derivatives of the electromagnetic 4-potential $F\mndd=\nabla_\mu A_\nu-\nabla_\nu A_\mu$; $\eta$ is the dimensionful coupling constant of the theory and $\mathfrak{f}\left(\phi\right)>0$ is a generic coupling function between the scalar field and the GB invariant $\rgb$, that is given by
\begin{equation}
\label{eq:rgb}
\rgb\equiv R^2-4R_{\mu\nu} R^{\mu\nu} +R_{\mu\nu\alpha\beta}R^{\mu\nu\alpha\beta}\,,
\end{equation}
with $R\left( R_{\mu\nu}\right)$ being the Ricci scalar (tensor) and $R_{\mu\nu\alpha\beta}$ the Riemann tensor. If not specified otherwise we use geometrized units in which $G=1=c$. Varying the action with respect to the dynamical fields one gets the coupled EMsGB system of equations,
\begin{align}
\label{eq:EMsGB_scalarKG}
\square\phi&=-\frac{\eta}{4} \frac{\partial \mathfrak{f}\left(\phi\right)}{\partial\phi}\rgb\,,\\
\label{eq:EMsGB_maxwell}
\nabla_\mu F^{\mu\nu}&=0\,,\\
\label{eq:EMsGB_einsteineq}
R\mndd-\frac{1}{2}g\mndd R&=T\mndd^{\left( F\right)}+\frac{1}{2} T\mndd^{\left(\Phi\right)}-\frac{1}{8}\eta\,T\mndd^{\left(\rgb\right)}\,,
\end{align}
where
\begin{align}
T\mndd^{\left(\phi\right)}&=\partial_\mu\phi\partial_\nu\phi-\frac{1}{2}g\mndd \partial_\alpha\phi\partial^\alpha\phi\,,\\
T\mndd^{\left( F\right)}&=2F_{\mu \alpha}F_\nu^{\,\,\alpha}-\frac{1}{2}F_{\alpha \beta}F^{\alpha\beta}g\mndd\,,\\
T\mndd^{\left(\rgb\right)}&= 16 R^\alpha_{(\mu}\mathcal{C}_{\nu)\beta} +8\mathcal{C}^{\alpha\beta}\left( R_{\mu\alpha\nu\beta}-g\mndd R_{\alpha\beta} \right) \nn\\
&\quad\,-8 \mathcal{C} G\mndd -4 R \mathcal{C}\mndd\,,
\end{align}
with 
\begin{equation}
\mathcal{C}\mndd= \nabla_\mu \nabla_\nu \mathfrak{f}\left(\phi\right)=\mathfrak{f}'\nabla_\mu\nabla_\nu\phi+\mathfrak{f}''\nabla_\mu\phi\nabla_\nu\phi\,,
\end{equation}
$\mathcal{C}=g^{\alpha\beta}\mathcal{C}_{\alpha\beta}$, ``prime" denotes derivative with respect to the argument and parentheses in the subscripts indicate symmetrization.

We shall be interested in models for which 
\begin{equation}
     \frac{\partial \mathfrak{f}}{\partial\phi}\left(\phi=0\right)=0 \ ,
     \label{zeroderi}
\end{equation}
such that $\phi=0$ together with any electrovacuum GR solution, is a solution of this model.
Moreover, if $\eta<0$ and
\begin{equation}
     \frac{\partial^2 \mathfrak{f}}{\partial\phi^2}\left(\phi=0\right)>0 \ ,
\end{equation}
tachyonic perturbations exist in regions where $\rgb<0$. This is the trigger of GB$^-$ scalarization. The simplest GB coupling function obeying these conditions, which is also the small $\phi$ approximation of a generic analytic coupling function obeying these conditions, is a quadratic function:
\begin{equation}
\label{eq:coupling_function}
\mathfrak{f}\left(\phi\right)=\phi^2/2\,.
\end{equation}
Indeed, in linear perturbation theory, Eq.\til\eqref{eq:EMsGB_scalarKG}, together with Eq.\til\eqref{eq:coupling_function}, takes the simple form,
\begin{equation}
\left(\square -\mu^2_{\rm eff}\right)\phi=0\,,
\end{equation}
where 
\begin{equation}
\label{eq:tachyonic_mass}
\mu^2_{\rm eff}=-\frac{\eta}{4}\rgb\,.
\end{equation}
When Eq.\til\eqref{eq:tachyonic_mass} takes negative values, the scalar field acquires a tachyonic mass. If there exists a region in which this condition is met, then
a tachyonic instability potentially occurs~\cite{Hod:2020jjy}. A simple calculation shows that for Kerr BH this happens if $j >     0.5$\til\cite{Dima:2020yac}. This triggers spin-induced (or GB$^-$) scalarization.

%%%%%%%%%%%%%%%%%%%%%%%%%%%%%%%%%%%%
\section{Magnetized solutions}
\label{Magnetized_solutions}
%%%%%%%%%%%%%%%%%%%%%%%%%%%%%%%%%%%%
Under the condition~\eqref{zeroderi}, the system\til\eqref{eq:EMsGB_scalarKG}-\eqref{eq:EMsGB_einsteineq} admits a consistent truncation corresponding to  the Einstein-Maxwell (EM) theory, together with  $\phi=0$. Thus electrovacuum GR solutions and a trivial scalar field are admissible solutions of the full set of equations of motion. We now describe the EM solutions describing BHs in magnetic Universes.

The non-singular Melvin Universe\til\cite{Melvin:1963qx} is an exact solution of EM theory that represents a cylindrically symmetric, non-singular, non-asymptotically flat, clump of self-gravitating magnetic flux lines in equilibrium. It is given by
\begin{equation}
\label{eq:Melvin_metric}
ds^2=\Lambda^2 \left(-dt^2+d\rho^2+dz^2\right)+\Lambda^{-2}\rho^2 d\varphi^2\,,
\end{equation}
written in cylindrical coordinates ($t,\rho,z,\varphi$).
The function $\Lambda$ introduces the strength of the magnetic field $B$, which is the only parameter describing the solution, and it only depends on the radial cylindrical coordinate:
\begin{equation}
\label{eq:Lambda_def_Melvin}
%\Lambda=1+\left(1/4\right)B^2\rho^2\,.
\Lambda=1+\frac{1}{4}B^2\rho^2\,.
\end{equation}
Many properties of this magnetic Universe and generalizations thereof have been studied over the years - see $e.g$~\cite{Gibbons:2001sx,Tseytlin:1994ei,Bambi:2015sla,Garfinkle:2011mp,Kastor:2020wsm,Gutperle:2001mb,Costa:2001ifa, Junior:2021dyw,Junior:2021svb}.

It is possible to add a neutral BH, with horizon mass $M$, inside the Melvin magnetic Universe described by Eq.\til\eqref{eq:Melvin_metric}. Remarkably, an exact 2-parameter solution to such problem exists, which reduces to the Schwarzschild spacetime for $B=0$, and to the Melvin magnetic Universe for $M=0$.  In usual spherical polar coordinates ($t,r,\theta,\varphi$), the metric of such a Schwarzschild-Melvin BH spacetime takes the simple form,
\begin{equation}
\label{eq:SMBH_metric}
ds^2=\Lambda^2(-f dt^2+f^{-1} dr^2+r^2d\theta^2)+\Lambda^{-2}r^2 \sin^2\theta d\varphi^2\,,
\end{equation}
where $f=1-2M/r$, with $M$ representing the mass of the spacetime 
\cite{Radu:2002pn} and $\Lambda$ is now given by
\begin{equation}
\label{eq:Lambda_def_SMBH}
%\Lambda=1+\left(1/4\right)B^2r^2\sin^2\theta\,.
\Lambda=1+\frac{1}{4}B^2r^2\sin^2\theta\,.
\end{equation}

Further increasing the complexity, it is also possible to add a charge parameter $Q$ to the previous BH in the Melvin magnetic Universe\til\cite{Ernst:1976mzr}. An exact 3-parameter solution to such problem exists, which reduces to the Reissner-Nordstr\"om (RN) solution for $B=0$, to the previous Schwarzschild-Melvin spacetime for $Q=0$ and to the Melvin magnetic Universe for $M=0=Q$\til\cite{Ernst:1976mzr}. The metric of such RN-Melvin BH is given by
\begin{align}
\label{eq:RNM_metric}
ds^2&=\lvert\Lambda\rvert^2(-\tilde{f} dt^2+\tilde{f}^{-1} dr^2+r^2d\theta^2)\nn\\
&\quad\,+\lvert\Lambda\rvert^{-1}r^2 \sin^2\theta \left(d\varphi-\tilde{\omega} dt\right)^2\,,
\end{align}
where $\tilde{f}=1-2M/r+Q^2/r^2$, with $Q$ being the electric charge of the BH and
\begin{align}
\label{eq:Lambda_def_RNMBH}
\Lambda&=1+\left(1/4\right)B^2\left(r^2\sin^2+Q^2\cos^2\theta\right)-i B Q \cos\theta\,,\\
\tilde{\omega}&=-2B Q r^{-1}+B^3 Q r +\left(1/2\right)B^3 Q^3 r^{-1}\nn\\
&\quad\,-\left(1/2\right)B^3 Q r^{-1}\left(r^2-2M r +Q^2\right)\sin^2\theta +{\rm cst}\,.
\end{align}

The novel feature of the RN-Melvin spacetime is that, due to the non-trivial Poynting vector, resulting from the existence of an electric charge inside a magnetic field, the spacetime becomes \textit{stationary}, rather than static, despite the $B=0$ limit being static. Loosely speaking, a RN BH in a Melvin Universe starts to spin. See, $e.g.$~\cite{Santos:2021nbf} for the full solution, including the gauge field.

It is also possible to add a Kerr, or Kerr-Newman (KN) BH inside the Melvin magnetic Universe, as long as the symmetry axes are  aligned. Such solution was first analysed in a linear setup by Wald\til\cite{Wald:1974np}. Soon after, the fully non-linear rotating BH solution containing back-reacting magnetic fields was determined in the seminal works of Ernst\til\cite{Ernst:1976mzr}, and Ernst and Wild\til\cite{doi:10.1063/1.522875}. Generalizations and extensions of Ref.\til\cite{Ernst:1976mzr} have been attained (see\til\cite{Budinova:2000yd,Bicak:2006hs} for detailed overviews and\til\cite{Gibbons:2013dna,Astorino:2016hls,Booth:2015nwa,Astorino:2015naa,Astorino:2016ybm} for details on their thermodynamic properties). 

The full metric of a KN-Melvin BH is displayed in detail in Ref.\til\cite{Gibbons:2013yq}, where it has been obtained from magnetizing a seed KN BH through solution-generating techniques, developed by Harrison\til\cite{doi:10.1063/1.1664508}. The resulting spacetime is an axially symmetric geometry that depends on the $\left\{r,\theta\right\}$ spacetime components. Furthermore, in general, it depends on the parameters of the seed KN BH considered before the magnetization procedure, $i.e.$ $\left\{M,J,q\right\}$ and on the magnetic field parameter $B$ introduced by the procedure. The (rather) lengthy geometry of a KN-Melvin BH it is given explicitly in Appendix B of Ref.\til\cite{Gibbons:2013yq}. For simplicity, we discuss here only its main properties without displaying all its components\footnote{In comparison with the original metric in\til\cite{Gibbons:2013yq}, we are setting the magnetic charge $p$ to zero. This further condition requires the azimuthal angle $\varphi$ to have period $\Delta\varphi=2\pi\left(1+\left(3/2\right)q^2 B^2 +2 a q M B^3 +\left(a^2M^2 +\left(1/16\right)q^4\right)B^4\right)$, in order to avoid conical deficit at the poles\til\cite{Hiscock:1980zf}.},
\begin{align}
\label{eq:KNM_metric}
ds^2&=H\left[-f dt^2+R^2\left(\frac{dr^2}{\Delta}+d\theta^2\right)\right]\nn\\
&\quad\,+\frac{\Sigma\sin^2\theta}{H R^2} \left(d\varphi-\omega dt\right)^2\,,\\
\label{eq:KNM_vector_potential}
A_\mu&=\left(\Phi_0-\omega \Phi_3,0,0,\Phi_3\right)\,,
\end{align}
where
\begin{align}
R^2&= r^2+a^2\cos^2\theta\nn\,\\
\Delta&= \left(r^2+a^2\right)-2M r +q^2\,,\nn\\
\Sigma&= \left(r^2+a^2\right)^2-a^2 \Delta\sin^2\theta\,,\nn\\
f&= R^2 \Delta \Sigma^{-1}\,,
\end{align}
where $a$ is defined through the usual KN BH angular momentum ($a \equiv J/M$) and $H,\omega,\Phi_0,\Phi_3$ are all complicated functions of $\left(r,\theta \right)$
and $\left(M,a,q,B \right)$, given, respectively, in Eqs. (B.5), (B.8), (B.15) and (B.17) of\til\cite{Gibbons:2013yq}. To ensure the validity of our findings, we have verified that the metric and vector potential in Eqs.\til\eqref{eq:KNM_metric}-\eqref{eq:KNM_vector_potential} satisfy the Einstein-Maxwell system and that correctly tend to the KN geometry when $B=0$.

The KN-Melvin geometry has for a long time been thought to tend asymptotically to the Melvin Universe. Instead, only relatively recently\til\cite{Gibbons:2013yq}, it has been shown that this is the case only if
\begin{equation}
\label{eq:ergoregion_condition}
q=-a M B\,.
\end{equation}
The condition in Eq.\til\eqref{eq:ergoregion_condition} ensures that the ergoregion surrounding the KN-Melvin BH does not extend to spatial infinity, and we are going to restrict to such case in the following analysis of the KN-Melvin, that we call \textit{Case I}. 

Despite the asymptotic property that distinguishes it, Case I has the caveat that the seed BH (in the magnetization procedure) is already electrically charged and one may object that this is not the ideal to try to describe an astrophysical system wherein the BH is expected to have negligible electric charge. Thus, we also consider the Kerr-Melvin solution as \textit{Case II} in which the seed BH has no electric charge, in which case $q=0$ and condition~\eqref{eq:ergoregion_condition} is not imposed. As we shall see below, the main conclusions will be common to both Case I and Case II.

In preparation for the next section, let us specify the horizons' positions in Eq.\til\eqref{eq:KNM_metric}. As usual, they are located at the roots of $1/g_{rr}$. Following the notation of\til\cite{Gibbons:2013yq}, and restricting to Case I, one can find that
\begin{equation}
r_{\pm}=\left(1 \pm \tilde{\varepsilon}\right)M\,,
\end{equation}
where $\tilde{\varepsilon}^2=1-\varepsilon^2$, and $\varepsilon$ is defined through,
\begin{equation}
\label{eq:varepsilon_def}
\varepsilon\equiv \frac{a}{M}\sqrt{1+B^2 M^2}\ ,\qquad {\rm with} \ \ \  0\leqslant \varepsilon \leqslant 1\,.
\end{equation}
From Eq.\til\eqref{eq:varepsilon_def} it is clear that if $B M \neq 0$, since $0\leqslant j=a/M \leqslant 1$, KN BH with a sufficiently large value of $j$ cannot support external magnetic fields without developing naked singularities, hence they are non-physical. For Case II instead, the event horizon location does not differ from the well-known Kerr case.

%%%%%%%%%%%%%%%%%%%%%%%%%%%%%%%%%%%%
\section{Scalarization and magnetized spacetimes}
\label{Scalarization_and_magnetized_spacetimes}
%%%%%%%%%%%%%%%%%%%%%%%%%%%%%%%%%%%%

In the previous section we have set the stage to analyse Case I and Case II of KN-Melvin spacetimes. Now, let us analyse the necessary condition that triggers the scalarization instability. 

As mentioned earlier, the effect of a self-gravitating magnetic field, \textit{per se}, could support the GB$^-$ instability. In a Schwarzschild-Melvin BH spacetime, near the horizon the GB invariant has the form,

{\small
\begin{align}
\label{eq:rgb_SMBH}
&\rgb\left(r\sim 2M\right)=\left({4 M^4 \left(\left(BM \sin \theta \right)^2 +1\right)^8}\right)^{-1} \times\nn\\
& \bigg[ 3 \left[\left(BM \sin\theta\right)^8  -2 \left(BM \sin\theta\right)^4 +\left(\left(BM\right)^4 \sin ^2\left(2 \theta \right)-1\right)^2\right]\nn\\
&\quad\;+24 \cos ^2\theta \left(\left(BM\right)^8 \sin ^6\theta -\left(BM\right)^6 \sin ^4\theta +\left(BM\right)^2\right)\nn\\
&\quad\;+16 \left(BM\right)^4 \cos ^4\theta  \left(1-6 \left(BM\right)^2 \sin ^2\theta \right)\bigg]\,.
\end{align}
}
Thus, there exists a range of angles $\theta$ such that $\rgb$ assumes negative values whenever, 
\begin{equation}
BM>0.971\,.
\end{equation}
Therefore, for sufficiently large $BM$ the necessary condition for scalarization (near the BH) is obeyed. For the Kerr BH the onset of the instability coincides with the emergence of a negative GB spacetime region (near the horizon). Thus the necessary condition for GB$^-$ scalarization is also sufficient.\footnote{Our goal will be to show that the necessary condition for scalarization in the KN-Melvin case is more stringent for $j$, when considering only the geometry near the horizon.} Assuming a similar scenario in the presence of the magnetic field, the above calculation provides us with an estimate for the onset of ${\rm GB}^-$ scalarization in static spacetimes; restoring physical units, we obtain,
\begin{equation}
B \simeq 2\times 10^{13} \, {\rm G} \left(\frac{4\times 10^6 M_{\odot}}{M}\right)\,,
\end{equation}
where we normalized to the mass of Sagittarius A$^*$. The above number above is relatively high, but still within the regime of the highest magnetic fields ever measured\til\cite{Olausen:2013bpa}.

\begin{figure}[h!]
\centering
\includegraphics[width=8.9cm,keepaspectratio]{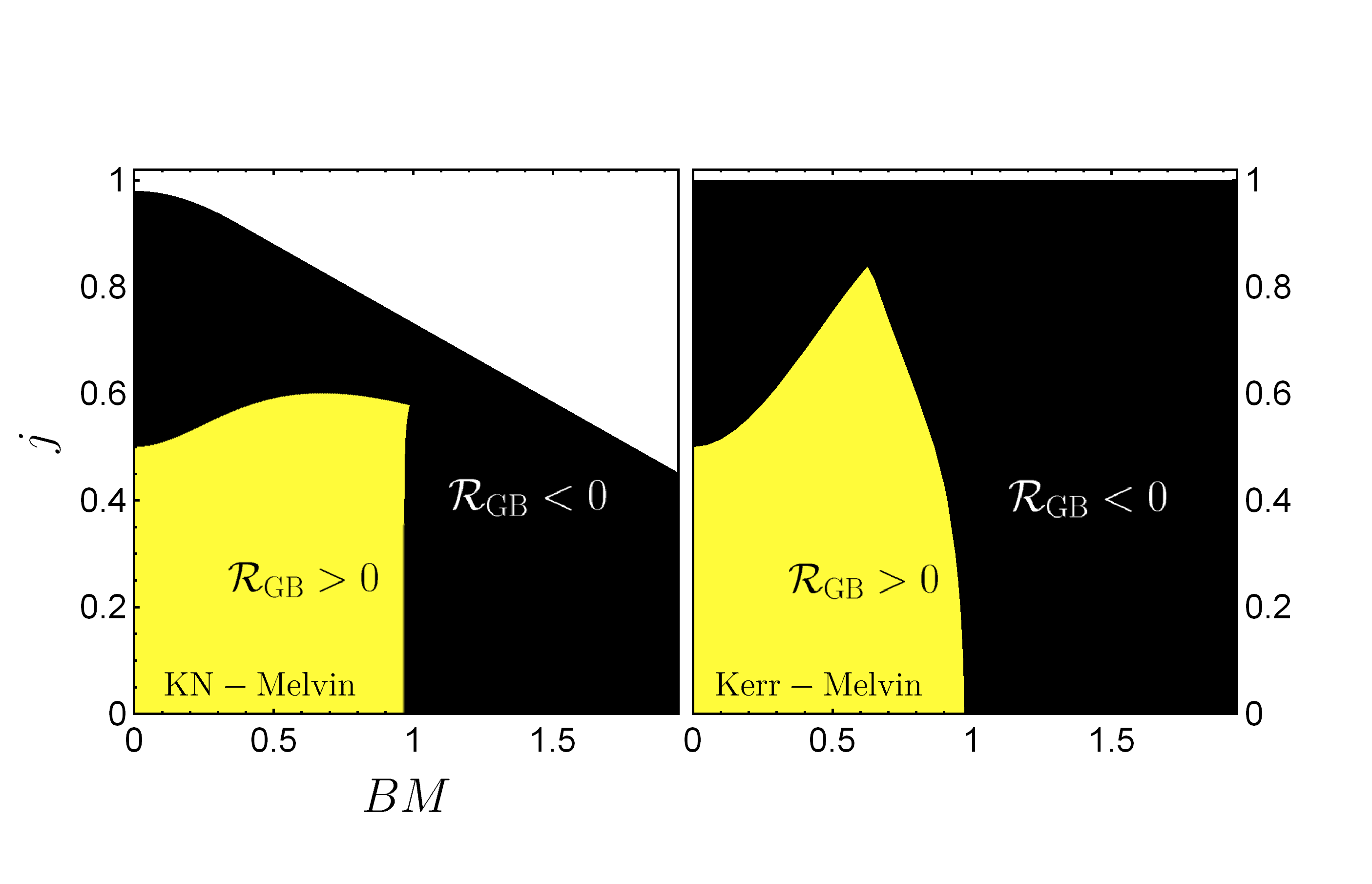}
\caption{Analysis of the sign of the GB invariant on the BH horizon for Case I (left panel) and Case II (right panel). The yellow region represents the domain in the $\left\{B M, j\right\}$ phase-space, where the GB invariant takes positive values for {\it any} azimuthal angle $\theta$ at the BH horizon. For black regions, on the other hand,  there is at least one value of $\theta$ such that $\rgb$ takes negative values. In the left panel, no KN-Melvin BHs are allowed in the top white region of the phase-space. That is due to the requirement of the existence of an event horizon for a KN-Melvin BH, $cf.$ eq.\til\eqref{eq:varepsilon_def}.}
\label{fig:KM_MphysBvsJphys}
\end{figure}

For what concerns KN-Melvin BHs, a similar analysis can be done. However, given the size of Eq.\til\eqref{eq:KNM_metric}, it is impractical to show the full form of $\rgb$ here. Therefore, in Fig.\til\ref{fig:KM_MphysBvsJphys} we display the regions where $\rgb$ becomes negative for at least one value of $\theta$, in a $\left\{B M ,j\equiv J/M^2\right\}$ phase-space, for both Case I (left) and Case II (right).

%%%%%%%%%%%%%%
%%%%%%%%%%%%%%%%%%

In the limit where $j\rightarrow 0$, we get the expected threshold for Schwarzschild-Melvin BHs (see Eq.\til\eqref{eq:rgb_SMBH}). Conversely, for small $B$, the threshold to have a negative $\rgb$ near the horizon is obtained for $j\geqslant 0.5$, as expected. However, starting from $B M=0,\,j=0.5$ and increasing the magnetic field, the necessary condition to have unstable static scalar perturbations is satisfied for larger values of $j$, compared to the Kerr BH limit. This perhaps surprising result shows that the effect of adding an external magnetic field to a Kerr BH is not making the scalarization process {\it easier} to happen, instead, it leads to GR solutions that are more stable than their non-magnetized counterparts, regardless of the case considered. 
A similar result is evident also considering slowly-rotating BHs in a neighborhood of the scalarization threshold of Schwarzschild-Melvin BH. Ultimately, Fig.\til\ref{fig:KM_MphysBvsJphys} indicates that rotation and magnetic fields possess different and somehow opposing roles \textit{near} the horizon, although they may both trigger the instability of KN-Melvin spacetimes separately. These results can be straightforwardly generalized to bosonic fields with non-zero spins.

An important remark is in order concerning physical quantities. The mass and angular momentum parameters in Fig.\til\ref{fig:KM_MphysBvsJphys} refers to the seed Kerr (or KN) BH ones. A definition of mass and angular momentum in spacetime with non-flat asymptotics quantities -- through Komar integrals, for instance -- may be difficult to obtain. Yet it has been shown\til\cite{Astorino:2016hls} that in the case of KN-Melvin BHs, it is possible to use the so-called canonical integrability method\til\cite{Barnich:2001jy,Barnich:2003xg,Barnich:2007bf} to obtain the physical quantities for such spacetimes. Following Ref.\til\cite{Astorino:2016hls}, we have checked that, using the physical BH parameters, the qualitative behaviour seen is similar to that in the left panel of Fig.\til\ref{fig:KM_MphysBvsJphys}; $i.e.$ it is consistent with using the seed parameters.

%%%%%%%%%%%%%%%%%%%%%%%%%%%%%%%%%%%%%%%%%
\subsection{A geometrical interpretation}
%%%%%%%%%%%%%%%%%%%%%%%%%%%%%%%%%%%%%%%%

As previously stated, the necessary condition to undergo tachyonic instability concerns the \textit{4-dimensional} $\rgb$ invariant. However, let us try to interpret the results of Fig.\til\ref{fig:KM_MphysBvsJphys} inspecting the horizon (2D) geometry of the KN-Melvin geometry.

Previous studies have shown that both rotation and the inclusion of magnetic fields deform the geometry of the BH external horizon\til\cite{Smarr:1973zz,Wild:1980zz,PhysRevD.33.2780,Costa:2009wj,Gibbons:2009qe,Junior:2021dyw}. Concretely, the effect of rotation turns the Kerr horizon into an oblate spheroid (compared to the spherical Schwarzschild horizon), perpendicularly to the angular momentum of the rotating BH ($z-$axis). The larger the rotation the more oblate becomes the horizon shape. In order to visualize the BH horizon, one might embed it isometrically in Euclidean 3-dimensional space. Notably, for Kerr BHs, such procedure is possible only for relatively slow BHs (precisely, up to $a=\sqrt{3}/2 M$).

\begin{figure}[ht]
\centering
\includegraphics[width=8.9cm,keepaspectratio]{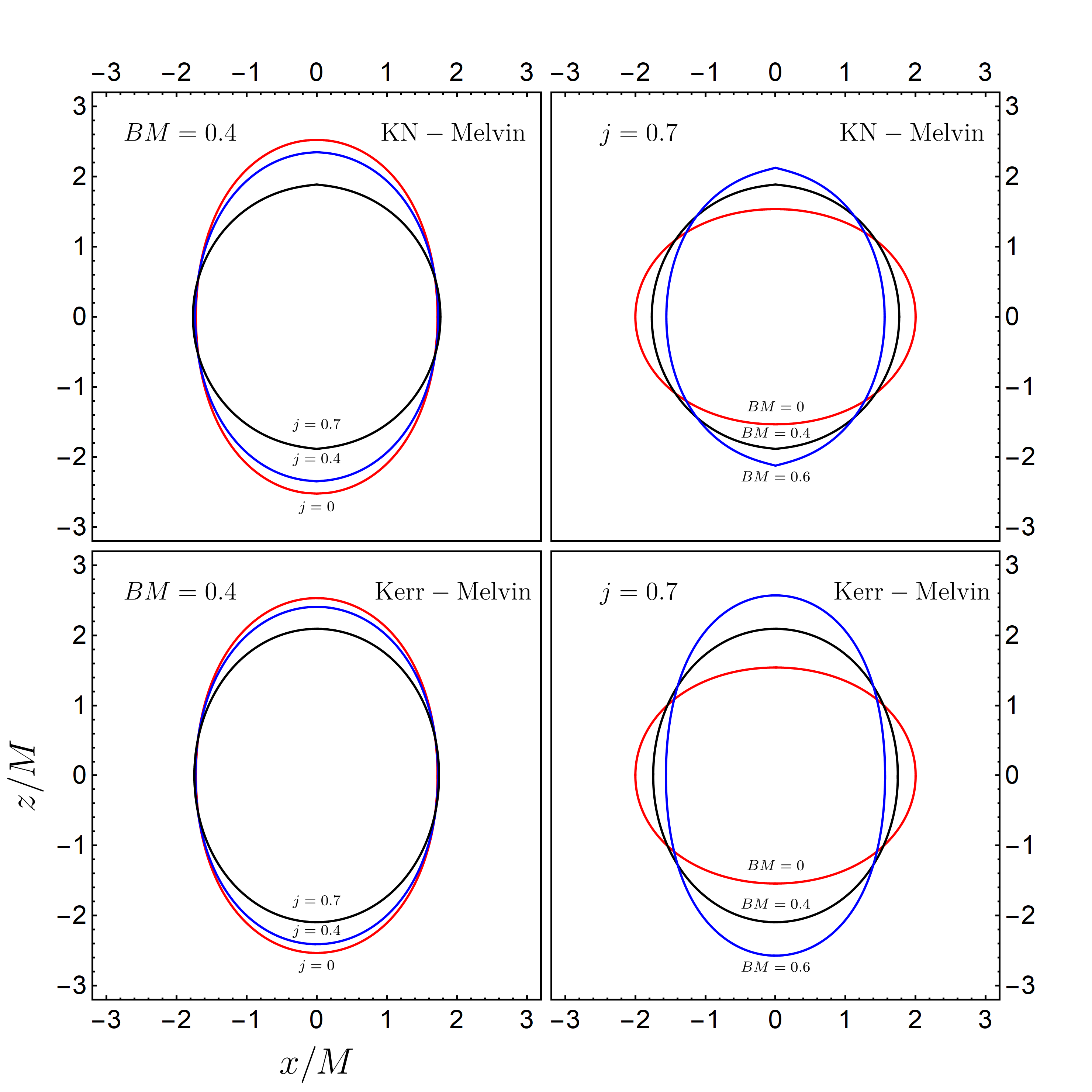} 
\caption{BH horizon isometric embeddings in Euclidean 3-dimensional space ($x-z$ sections)  for different values of $\left\{B M,j\right\}$. As rotation increases, the horizon becomes oblate (perpendicularly to the axis of rotation). Increasing $B$ instead elongates the horizon along the $z-$direction (prolate). In the case where both rotating and magnetic fields are present, the structure resembles the static case.
}
\label{fig:plot_embedding}
\end{figure}

Likewise, the horizon structure of a Schwarzschild-Melvin BH is deformed in comparison to the static BH case. However, conversely to the Kerr case, the horizon geometry is deformed in the same direction as the external magnetic field, making it a prolate spheroid. As expected, a larger $B M$ corresponds to a more deformed horizon structure. In order to visualize the horizon structure of KN-Melvin BHs, we followed the work of Smarr\til\cite{Smarr:1973zz}, but including an external magnetic field. Our results are shown in Fig.\til\ref{fig:plot_embedding}: the inclusion of magnetic fields counter-balance the effects of rotation, yielding a more spherical horizon shape to spinning and magnetized spacetimes. This provides some insight on how the rotation and magnetic field have a conflicting geometric effect, once a KN BH is placed in the external magnetic field. 

%%%%%%%%%%%%%%
%%%%%%%%%%%%%%%%%%

%%%%%%%%%%%%%%%%%%%%%%%%%%%%%%%%%%%%
\section{Conclusions}
\label{sec:conclusions}
%%%%%%%%%%%%%%%%%%%%%%%%%%%%%%%%%%%%
%In the next decades, GWs interferometers\til\cite{Hild:2010id,LISA:2017pwj,Audley:2017drz,Punturo:2010zza,Maggiore:2019uih,10.1093/nsr/nwx116} should be able to test the effects of matter around compact objects, of which electromagnetic fields are an example. 

%GWs from compact sources might be as well used to test modifications to Einstein's equations of motion.  
%signals directly generated in highly curved spacetime regimes (compared to the Solar system surroundings for instance) 
%Among the latter, {\it smoking guns} for new physics also include finding solutions of the chosen model that may represent new type of compact objects, which differ from their GR counterpart. 
%The latter can be achieved, for instance, including hypothetical new fields or particles around isolated objects or binary systems\til\cite{Berti:2015itd,Barack:2018yly,Barausse:2020rsu,Cardoso:2011xi,Brito:2015oca,Herdeiro:2014goa,Herdeiro:2015waa}. 
%As we will shortly see, the presence of extra parameters needed to describe BH spacetimes might spoil the universality of the KN family. The introduction of new fundamental scales indeed allows for richer BH solutions, whose geometry in general will depend on the ratio between their charges ($M, Q,$ etc.) and the extra coupling constant of the chosen theory.

A paradigmatic mechanism that leads to new field configurations is given by ``spontaneous scalarization''. This process has been discovered in the framework of scalar-tensor theories\til\cite{Damour:1993hw}, where a scalar field, coupled to gravity, triggers a tachyonic instability, leading to stars with non-trivial scalar charges. These compact stars are said to be {\it scalarized}~\cite{Damour:1992we}\footnote{Spontaneous growth of fields associated with tachyonic instabilities may also occur for vectors, tensors and spinors\til\cite{Ramazanoglu:2017xbl,Doneva:2017duq,Annulli:2019fzq,Kase:2020yhw,Ramazanoglu:2019gbz,Ramazanoglu:2017yun,Ramazanoglu:2018hwk,Minamitsuji:2020hpl,Blazquez-Salcedo:2020nhs,Herdeiro:2020htm}. }. A similar scalarization mechanism might also happen in vacuum BH spacetimes, when non-trivial couplings between a scalar and (higher-order) curvatures are present\til\cite{Silva:2017uqg,Doneva:2017bvd,Witek:2018dmd,Silva:2018qhn,Minamitsuji:2018xde,Doneva:2019vuh,Fernandes:2019rez,Minamitsuji:2019iwp,Cunha:2019dwb,Andreou:2019ikc,Ikeda:2019okp,Annulli:2021lmn}. In these theories, scalarized BHs differ from the Kerr family because some of the assumption of the above-mentioned no-hair theorems cease to be valid. 
%From an observational perspective, GW observations might directly place constraints on additional BH charges, if, and when, non-GR numerical methods will accurately model GWs generation in alternative theories of gravity (see Ref.\til\cite{East:2020hgw} as an example).

Recent works in the context of Einstein-scalar-Gauss-Bonnet theory have shown that both static and fastly spinning BHs might suffer from tachyonic instabilities\til\cite{Silva:2017uqg,Doneva:2017bvd,Dima:2020yac}, depending on the sign of the GB curvature invariant ($\rgb$).  Further details on the properties of rotating and non-rotating scalarized BHs in EsGB have been obtained, among others, in Refs.\til\cite{Blazquez-Salcedo:2018jnn,Collodel:2019kkx,Berti:2020kgk,Pierini:2021jxd}. ${\rm GB}^-$ scalarization is not a unique feature of rotating solutions, as recently shown by Ref.\til\cite{Herdeiro:2021vjo} for charged BH in EMsGB theory. In fact, even object {\it solely} formed by magnetic fields, as the Melvin Universe\til\cite{Melvin:1963qx}, might lead to {\it magnetic-induced} scalarization\til\cite{Brihaye:2021jop}.

Magnetic fields are ubiquitous in nature. Their impact on BHs, and particles around them, can be crucial in the current GW astronomy era. There are in fact known stars which encompass magnetic fields up to $10^{16} {\rm G}$\til\cite{Olausen:2013bpa}. The interaction of such magnetars and BHs might lead to unexpected and intriguing new phenomena. Simultaneously, it is fundamental as well to enlarge our predictive power, searching for extra BH charges coming from modification of GR. Hence, in the context of EMsGB theories, in this paper we have shown that astrophysical BHs tend to be more stable, once magnetic fields are added to Kerr spacetimes. The two effects of rotation and magnetic field are therefore substantially different in the near horizon region of KN-Melvin BHs, at least up to $BM\sim 1$.

A thorough analysis of scalar perturbations in such theory might be possible. Practically this means solving for static scalar perturbations around a KN-Melvin BH. This study would provide the exact values for the onset of the instability in the $B M$ and $j$ phase-space. However, going through this path raises important issues. In fact, as previously mentioned, a cutoff in the magnetic field would be needed in order to ensure asymptotic flatness, and this procedure introduces ambiguities. Furthermore, the results exhibited in Fig.\til\ref{fig:KM_MphysBvsJphys} clearly show that adding a magnetic field is not enhancing the scalarization process of KN BH thus making the result in this paper clear: spin-induced scalarization is \textit{not} facilitated by magnetic fields, at least poloidal ones that can be approximately described, near the horizon, by the Melvin magnetic Universe, as long as $BM\lesssim 1$.

%As a final note, despite their appealing properties, the non-flatness asymptotics of magnetized BH spacetimes makes these solutions not viable to describe isolated objects in nature. However, their near horizon geometry is thought to approximately describe the surroundings of astrophysical BHs immersed in external magnetic fields. Therefore, we may take magnetized BH solutions as toy-models for BHs immersed in such environments.

%%%%%%%%%%%%%%%%%%%%%%%%%%%%%%%%%%%%
\subsection*{Acknowledgements}
%%%%%%%%%%%%%%%%%%%%%%%%%%%%%%%%%%%%
We thank Haroldo Lima Junior for important feedback. This work is supported by the Center for Research and Development in Mathematics and Applications (CIDMA) through the Portuguese Foundation for Science and Technology (FCT - Fundaç\~{a}o para a Ci\^{e}ncia e a Tecnologia), references UIDB/04106/2020, UIDP/04106/2020. LA is supported by the University of Aveiro through a PostDoc research grant, reference BIPD/UI97/9854/2021. We acknowledge support from the projects PTDC/FIS-OUT/28407/2017, CERN/FISPAR/0027/2019 and PTDC/FIS-AST/3041/2020. This work has further been supported by the European Union’s Horizon 2020 research and innovation (RISE) programme H2020-MSCA-RISE-2017 Grant No. FunFiCO-777740.

%

%
%%%%%%%%%%%%%%%%%%%%%%%%%%%%%%%%%%%%%%%%%%%%%%%%%%%%%%%%%%%%%%%%%%%%%%%%%%%%%

%\clearpage
%\newpage
%\appendix

%%%%%%%%%%%%%%%%%%%%%%%%%%%%%%%%%%%%%%%%%%%%%%%%%%%%%%%%%%%%%%%%%%%%%%%%%%%%%%%%%%%%%%%%%%%%%%%%%%%%%%%%
%\section{Appendix A} \label{app:appendix_A}
%%%%%%%%%%%%%%%%%%%%%%%%%%%%%%%%%%%%%%%%%%%%%%%%%%%%%%%%%%%%%%%%%%%%%%%%%%%%%%%%%%%%%%%%%%%%%%%%%%%%%%%%

\bibliographystyle{apsrev4}

\bibliography{References}

\end{document}